\documentstyle[12pt]{article}
\title{Classical and quantum mechanics on information spaces
with applications to cognitive, psychological, social and 
anomalous phenomena}
\author{Andrei Khrennikov\\
Department of Mathematics, Statistics and Computer Sciences\\
University of V\"ax\"o, S-35195, Sweden}

\begin{document}
\maketitle

\footnote{This research was supported by the grant "Strategical Investigations"
of the University of V\"axj\"o and the visiting professor fellowship at the
University of Clermont-Ferrand.}

\begin{abstract} 
We use the system of $p$-adic numbers for the description of information processes.
Basic objects of our models are so called transformers of information,
basic processes are information processes, the statistics are information statistics
(thus we present a model of information reality).
The classical and quantum mechanical formalisms 
on information $p$-adic spaces are developed. 
It seems that classical and quantum mechanical models on $p$-adic information spaces
can be applied for the investigation of flows of information in cognitive and social
systems, since a $p$-adic metric gives quite natural description of the  
ability to form associations.
\end{abstract}

\section{Introduction}

We develop classical and quantum formalisms on {\it information spaces.}
Basic objects of this model are so called 
{\it transformers of information}; basic processes are 
information processes. Our main aim is a description of classical and quantum
dynamics of information states.

This information dynamics may be fruitful in the study of cognitive,
psychological and social processes. Here flows of information are more important
than flows of matter. We think that it would be possible to explain some
aspects of the process of thinking and psychological, social and anomalous
phenomena on the basis
of our model. Thus the readers who are only interested in applications
to cognitive sciences, sociology and psychology may consider our model
as only a new apparatus to investigate these phenomena. 

Our model of information reality can be considered as an attempt to extend
the standard model of physical reality.
We interpret material objects as
a particular class of transformers of information (which are characterized
by stable or slowly changing information states). On the other hand, our model
might be used for the description of information flows which are not directly related 
to flows of matter. These are conscious, social (and even anomalous) information
processes.

Different models of information and cognitive reality 
have been discussed by many scientists
in relation  to foundations of quantum physics [1]-[7],
cognitive sciences and psychology [8]- [11] and anomalous
phenomena [12]-[17].

We use a new mathematical apparatus to describe information reality
("the world of ideas").  Many authors discuss the idea
that such "ideal" objects as ideas, consciousness, information
can play an important role to provide the right picture of physical reality.
However, typically they use (with some modifications)  the standard mathematical 
model based on the description of physical reality by {\it real numbers.}
In particular, many of them discuss a "conscious field", but they 
try to describe  this object as a new field on the standard
real space-time. We think that some of cognitive processes could not be described
by using the real model of physical reality. There is simply no
place for such phenomena in this model.  The real model 
was created to describe a particular class of physical phenomena
(material objects). This model does not play an exceptional role. We need not try to input all
physical phenomena into this real model of reality. There can be other models of physical 
reality. We propose to describe physical reality by using
information spaces (see Appendix 1).

From our viewpoint real spaces (Newton's absolute space or spaces of general relativity)
give only a particular class of information spaces.  These real information spaces are characterized
by the special  system for the coding of information and the special distance on the space of
vectors of information. Any natural number $m > 1$ can be chosen as the basis of the coding system. 
Each $ x \in [0,1]$ can be presented in the form:
\begin{equation}
\label{RP}
x= a_0 a_1... a_n...\; ,
\end{equation} 
where $a_j= 1,...,m-1,$ are digits. We denote the set of all
sequences of the form (\ref{RP}) by the symbol $X_m.$
For example, let us fix $m= 10.$ One of the main properties of the real cording 
system is the identification of the form:
\begin{equation}
\label{ID}
10...0...= 09...9...\; ; 010....0...= 009...9...\; ;...
\end{equation}
In fact, this identification is closely connected with {\bf the order structure} 
on the real line ${\bf R}$ (and the metric related to this order structure).
 For each $x$, there exist "right" and "left"
 neighborhoods; there exist
arbitrary small right and left shifts. The identification (\ref{ID}) is connected with the 
description of left neighborhoods.

{\bf Example 1.1.} Let $x= 10...0... \; .$ Then $x$ can be approximated from the left
hand side with an arbitrary  precision by numbers of the form $y=09...90... \;.$

The following description of
right neighborhoods will be very important in our further considerations. 

($AS$) Let $x= a_0...a_m...\; .$ Then the numbers (vectors of information) which are close
to the $x$ from the right hand side  have the form $y=b_0...b_m...,$ where
$a_0=b_0,..., a_m=b_m$ for sufficiently large $m.$

This nearness has a natural information interpretation: ($AS$) implies the ability to form
associations for cognitive systems which use this nearness to compare vectors of information.
By ($AS$) two communications (two ideas in a model of human thinking, [18] - [20]) which have 
the same codes for sufficiently large number of first (the most important) positions in cording 
sequences are identified by a comparator of a cognitive system.

Numbers (vectors of information) which are close to $x$ from the left hand side could not be 
characterized in the same way (see Example 1.1, there $x$ and $y$ are very close but their codes
differ strongly). 

{\bf Conclusion.} The system of real numbers has been created as a coding system for information
which the consciousness receives from reality. The main properties of this coding system
are {\it the order structure} on the set of information vectors\footnote{Of course, the idea about 
an order structure is a consequence of properties of the special system which is used 
for observations of reality.}
and the restricted ability (see ($AS$) ) to form {\it associations.}

Finally, we pay attention to the "universal coding property" of the real system:
any natural number $m>1$ can be used as the basis of this system. Thus it is assumed
that any information process can be equivalently described by using, for example,
2-bits coding or 1997-bits coding.

All these properties of the real coding system were incorporated in every physical model
\footnote{From this point of view there is no large difference between Newton's absolute
space and real manifolds used in general relativity.}.

I do not think that all information processes have an order structure. On the other hand, 
the scale of coding system $m>1$ may play an important role in a description
of an information process.

Let us "modify" the real coding system. We eliminate the identification (\ref{ID}).
Since now, there is {\bf no order structure} on the set $X_m$ of information vectors. We consider
on $X_m$ the nearness defined by ($AS$)\footnote{Thus here all information is considered from the
viewpoint of associations.}. This nearness can be described by a metric. The corresponding
(complete) metric space is isomorphic to the ring of so called $m$-adic integers
${\bf Z}_m$ (see [21] and section 2). Therefore it is natural to use $m$-adic numbers for
a description of information processes. Mathematically it is convenient to use 
prime numbers $m = p > 1$ (see [21]). We arrive to the domain of an extended 
mathematical formalism, $p$-adic analysis. 

To use $p$-adic numbers in physics is not a new idea (see [22] - [40]). A new idea is to use them 
for a description of information (in particular, cognitive [18]) processes. On the other hand, 
apparatus which has been developed in $p$-adic quantum physics
may be fruitfully used in our model. 

We develop a quantum formalism for information systems. The 
mathematical basis for this formalism has been presented in  [29], [18], [37], [34], [35].
In this paper we apply the $p$-adic quantum formalism to information
systems. As in ordinary quantum mechanics over the reals, the problem of an interpretation
plays the important role in information quantum mechanics. Of course, all difficulties
of an interpretation of the ordinary quantum theory (see, for example, [6], [41] -- [45])
are reproduced in the information quantum theory. There are many viewpoints on 
an interpretation of the quantum theory (which may be very different). However, they
are mainly based on the following two general interpretations of a quantum state:
(1) an {\it individual (or orthodox Copenhagen) interpretation} by which a quantum state
provides the complete description an individual quantum system; (2) an {\it ensemble
(or statistical) interpretation} by which a quantum state provides the description
of a statistical ensemble of quantum systems. In fact, by analysing the process of measurement
for information quantum systems we understood that we have to follow the ensemble 
interpretation. This analysis might be also useful for better understanding of the ordinary
quantum formalism on real space.

\section{Systems of $p$-adic numbers}

First we present some facts about $p$-adic numbers.

The field of real numbers ${\bf R}$ is constructed as the completion of the
field of rational
numbers ${\bf Q}$ with respect to the metric $\rho(x,y)$ $=$ $\vert x - y
\vert$, where $\vert
\cdot\vert$ is the usual valuation  given by the absolute value.
The fields of $p$-adic
numbers ${\bf Q}_p$  are constructed in a corresponding way, but  using
other  valuations.
For a prime number $p$, the $p$-adic valuation $\vert \cdot\vert_p $ is
defined in the following
way. First we define it for natural numbers. Every natural number $n$ can
be represented as the
product of prime  numbers, $n$ $=$ $2^{r_2}3^{r_3} \cdots p^{r_p} \cdots$,
and we define
$\vert n\vert_p$ $=$ $p^{-r_p}$, writing $\vert 0 \vert_p$ $=0$  and
$\vert -n\vert_p$ $=$
$\vert n \vert_p$.  We then extend the definition of the $p$-adic valuation
$\vert\cdot\vert_p$
to all rational numbers by setting $\vert n/m\vert_p$ $=$ $\vert
n\vert_p/\vert m\vert_p$
for $m$ $\not=$ $0$. The completion of ${\bf Q}$ with respect to the metric
$\rho_p (x,y)$ $=$
$\vert x- y\vert_p$ is the locally compact field of $p$-adic numbers ${\bf
Q}_p$. The number fields ${\bf R}$ and 
${\bf Q}_p$ are unique in a sense, since by
Ostrovsky's
theorem (see [21]) $\vert\cdot\vert$ and  $\vert\cdot\vert_p$ are the
only possible  valuations
on ${\bf Q}$, but have quite distinctive properties. 

Unlike the absolute value distance $\vert \cdot \vert$, the $p$-adic
valuation satisfies the strong
triangle inequality
$|x+y|_p \leq  \max[|x|_p,|y|_p],  \quad x,y \in {\bf Q}_p$

Write $U_r(a)$ $=$ $\{x\in {\bf Q}_p: |x -a|_p \leq r\}$ and
 $U_r^-(a)$ $=$ $\{x\in {\bf Q}_p: |x -a|_p < r\},$
where $r$ $=$ $p^n$ and $n$ $=$ $0$, $\pm 1$, $\pm 2$,
$\ldots$.
These are the ``closed'' and ``open''  balls in ${\bf Q}_p$ 
while the sets
$S _r(a)$ $=$ $\{x \in K: |x -a|_p = r \}$ are the spheres in ${\bf Q}_p$
of such radii $r$.
These sets (balls and spheres) have a somewhat strange topological
structure from the
viewpoint of our usual Euclidean intuition: they are  both open and closed at
the same time, and as such are called {\it clopen} sets.  Another
interesting property of $p$-adic
balls is that  two balls have nonempty intersection if and only if one of
them is contained
in the other. Also, we note that any point of  a $p$-adic ball can be
chosen as its center,
so such a ball is thus not uniquely characterized by its center and radius.
Finally, any $p$-adic
ball $U_r(0)$ is an additive subgroup of ${\bf Q}_p$, while the ball
$U_1(0)$ is also
a ring, which is called the {\it ring of $p$-adic integers} and is denoted
by ${\bf Z}_p$.

Any $x$ $\in$ ${\bf Q}_p$ has a unique  canonical expansion (which
converges in the
$\vert \cdot\vert_p$--norm) of the form
$
x= a_{-n}/p^n +\cdots\ a_{0}+\cdots+ a_k p^k+\cdots
$
where the $a_j$ $\in$ $\{0,1,\ldots, p-1\}$ are the ``digits'' of the
$p$-adic expansion. The
elements $x$ $\in$ ${\bf Z}_p$ have the expansion
$
x=a_{0}+\cdots+ a_k p^k+\cdots
$
and can thus be identified with the sequences of digits
$
x = a_0 ... a_k ... .
$

The $p$-adic exponential function $e^x=\sum_{n=0}^{\infty}\frac{x^n}{n!}.$ The series
converges in $\mathbf{Q}_p$ if 
\begin{equation}
\label{EXP}
|x|_p\leq r_p , \; \mbox{where}\; r_p=1/p,\ p\ne 2 \; \mbox{and} \; r_2=1/4.
\end{equation}
$p$-adic  trigonometric functions $\sin x$ and $\cos x$ are defined by the standard
power series. These series have the same radius of convergence $r_p$ as the exponential
series.

If, instead of a prime number $p$, we start with an arbitrary natural number
$m > 1$ we construct the system of so called $m$-adic numbers ${\bf Q}_m$
by completing ${\bf Q}$ with respect to the $m$-adic metric $\rho_m(x,y)$ $=$
$|x-y|_m$ which is defined in a similar way to above.  However, this system
is in general not a field as there may exist divisors of zero.

\section{Dynamics on information spaces}

The rings of $p$-adic integers ${\bf Z}_{p}$ can be used as mathematical models for
information spaces. Each element $x=\sum_{j=0}^{\infty}\alpha_{j}p^{j}$ 
can be identified with a sequence
\begin{equation}
	x=\alpha_{0}\alpha_{1}\cdots\alpha_{N}\cdots ,\ \alpha_{j}=0,1,\ldots,p-1.
\end{equation}
 Such sequences are interpreted as coding
sequences (in the alphabet $A_{p}=\{0,1,\ldots,p-1\}$ with $p$ letters) for some amounts
of information. The $p$-adic metric $\rho_{p}(x,y)=|x-y|_{p}$ on ${\bf Z}_{p}$
corresponds to the  nearness  ($AS$) for information sequences.
We choose the space $X = {\bf Z}_p$ (or multidimensional spaces  $X = {\bf Z}_p^N$)
for the
description of information. The $X$ is said to be {\it information
space.}

Everywhere below we shall use the abbreviation $"I"$ for the word information
(for example, information space = $I$-space).

{\bf Remark 3.1.}  Different information
phenomena can be described by different mathematical models for $I$-spaces. The $p$-adic
model for $I$-spaces is the simplest from the mathematical point of view. 

Objects which "live" in $I$-spaces are said to be {\it transformers of information}
($I$-transformers). $I$-transformers are not characterized by localization in information
$p$-adic space (or real space). They are characterized by the ability to receive an external information
and transform it in a new information.

Each $I$-transformer $\tau$ has internal clocks. A state of the
clocks is described by an $I$-vector  $t\in T={\bf Z}_p$ which is called {\it information time}. 
The $I$-time can have different interpretations in different $I$-models. If $\tau$
is a conscious system then $t$ is (self-recognized) time of the evolution of this system.
We can say about psychological time of an individual or about
(collective) social time of a group of individuals. In fact, we have not to image
$t$ as an ordered sequence of time counts. This is only information with describes
evolution of $\tau.$ In principle, there is no direct relation between $I$-time and 
"physical" time that is used in the model over the reals.

At each instant $t\in T$ of $I$-time there is defined
a {\it total information state} ($I$-state) $q(t)\in X$ of $\tau.$ 
It describes the position of $\tau$ in the  $I$-space $X$. The
"life"-trajectory of $\tau$ can be identified with the trajectory $q(t)$ in $X$.

{\small An $I$-transformer can be imagine as a kind of Turing machine. Let us consider
a free $I$-transformer $\tau_{\rm{fr}}$ (i.e., an $I$-transformer which does not 
interact with other $I$-transformers and $I$-fields). At the instant of $I$-time
$t$ the $\tau_{\rm{fr}}$ has the $I$-state $q(t)= (\alpha_0(t), \alpha_1(t),...,
\alpha_k(t),...)$ (an infinite ribbon with symbols belonging to the alphabet 
$A_p =\{ 0,1,..., p-1\}).$ During an $I$-time interval $\Delta t$ this state is 
transformed in a new state $q(t+ \Delta t)= (\alpha_0(t+\Delta t), \alpha_1(t+\Delta t),...,
\alpha_k(t+\Delta t),...)$ (a new infinite ribbon with symbols belonging to the alphabet 
$A_p).$ The law of transformation depends on internal $I$-parameters $s$ 
which determine the internal structure of $\tau_{\rm{fr}}.$ In the general case an 
$I$-transformer $\tau$ interact with other $I$-transformers $\tau_j, j=1,...,N$
and $I$-fields $\phi_i(x), i=1,...,M.$ These interactions change continuously 
internal $I$-parameters $s=s(t, q_{\tau_j}(t)),\phi_i(q_\tau(t))).$

For example, a cognitive system $\tau$ which is isolated from external
$I$-flows can be considered as a free $I$-transformer $\tau_{\rm{fr}}.$
Here $q(t)$ gives the evolution of $\tau_{\rm{fr}}$ in `space of ideas';
$I$-parameters $s$  are determined by the neural structure of $\tau_{\rm{fr}}.$
In general case the cognitive system $\tau_{\rm{fr}}$ interact with other
cognitive systems and material objects (the latter interactions are also considered
by $\tau_{\rm{fr}}$ as $I$-interactions) and $I$-fields. These interactions change
continuously (with respect to $I$-time of $\tau_{\rm{fr}})$ the transformation law,
$q(t)  \to q(t+\Delta t).$

We consider now the motion of a material particle $\tau$ from the $I$-viewpoint.
At the moment we restrict our consideration to classical one dimensional motions.
We identify the total $I$-state $q$ of a particle $\tau$ with the spatial coordinate
of this particle. $q \in {\bf Z}_p$ has the form $q= \alpha_0 + \alpha_1 p +\cdots+
\alpha_m p^m +\cdots.$ This representation can be considered as the
expansion of the distance $q$ in the $p$-scale. The main difference from the real
model of the motion of $\tau$ is discreetness of space. There is the minimal length
element $l=1.$ The particle $\tau$ could not be observed on distances which are less
than $l=1.$ Other difference is that $q$ can yield infinitely large values
(these are $q$ for which $\alpha_j\not=0$ for an infinite number of $j).$ Thus
the realization of $I$-space as spatial space does not reproduce the ordinary 
model of motion in continuous real space. It gives a model of motion in discrete 
space. The ordinary physical interactions can realized in this space (see 
[18], [29], [34]-[36]). In this way they can be interpreted as $I$-interactions.}

We develop an analogue of the Hamiltonian dynamics on the $I$-spaces
\footnote{In fact, this is an application to the $I$-theory of the Hamiltonian
$p$-adic formalism developed in [26] (and generalized in [29]).}
 As usual, we
introduce the quantity $p(t)=\dot{q}(t)\  (=\frac{d}{dt}q(t))$ which is the information
analogue of the momentum. However, here we prefer to use a physiological terminology.
The quantity $p(t)$ is said to be a {\it motivation} (for changing of the $I$-state $q(t)$).

The space ${\bf Z}_{p}\times {\bf Z}_{p}$ of points $z=(q,p)$ where $q$ is the $I$-state and $p$ is
the motivation is said to be a phase $I$-space. As in the ordinary Hamiltonian formalism,
we assume that there exists a function $H(q,p)$ ($I$-Hamiltonian) on the phase $I$-space
which determines the motion of $\tau$ in the phase $I$-space:
\begin{equation}\label{l1}
	\dot{q}(t)=\frac{\partial H}{\partial p}(q(t),p(t)),\ q(t_0)=q_0,
\end{equation}
\begin{equation}\label{l2}
	\dot{p}(t)=-\frac{\partial H}{\partial q}(q(t),p(t)),\ p(t_0)=p_0.
\end{equation}
The $I$-Hamiltonian $H(p,q)$ has the meaning of an $I$-{\it energy.} In principle, $I$-energy is
not related to the usual physical energy.

{\small If $\tau$ is a (material) particle, then (\ref{l1}), (\ref{l2}) gives the
Hamiltonian dynamics for the particle; here $q(t)$ is the spatial coordinate of
the particle in discrete space and $p(t)$ is the momentum of the particle (which
is also discrete). If $\tau$ is a cognitive system, then (\ref{l1}), 
(\ref{l2}) gives the Hamiltonian dynamics for the cognitive system in the `space 
of ideas'.}

The simplest $I$-Hamiltonian $H_{\rm{fr}}(p)=\alpha p^2,\ \alpha\in Z_p$ describes the
motion of a free $I$-transformation $\tau$, i.e., an $I$-transformer which uses only
self-motivations for changing of its $I$-state $q(t)$. Here by solving the system of the
Hamiltonian equations we obtain: $p(t)=p_0,\ q(t)=q_0+2\alpha p_{0}(t-t_0)$
\footnote{In fact, this simplest $I$-system is not trivial from the mathematical
viewpoint. There exist other solutions which are nonanalytic (but smooth), see [21], [46].
These solutions may also have an interesting $I$-interpretation. We shall
discuss this problem later.}.
The motivation $p$ is the constant of this motion. Thus the free $I$-transformer
"does not like" to change its motivation $p_0$ in the process of the motion in the
$I$-space. If, we change coordinates, $q'=(q-q_0)/k,\ k=2\alpha p_0$, then we see that the
dynamics of the free $I$-transformer coincides with the dynamics of its $I$-time.

{\small If $\tau$ is a (material) particle, then $p_0$ is its momentum and 
$\alpha= 1/2m,$ where $m (=m_0 +m_1 p+\cdots+m_l p^l)$ is the mass of $\tau$ 
(which determined with a finite precision). If $\tau$ is a cognitive system,
then $p_0$ is (internal) motivation of $\tau$ and $\alpha= 1/2m,$ where
$m$ is so called $I$-mass (see section 4).}

In general case the $I$-energy is the sum of the $I$-energy of motivations
$H_{f}=\alpha p^2$ (which is an analogue of the kinetic energy) and potential $I$-energy
$V(q)$:
$$H(q,p)=\alpha p^2+V(q).$$
The potential $V(q)$ is determined by {\it fields of information}. 

In the Hamiltonian framework 
we can consider interactions between  $I$-transformers
$\tau_1,\ldots,\tau_N$. These $I$-transformers have the $I$-times $t_1,\ldots,t_N$ and 
$I$-states $q_1(t_1),\ldots,q_N(t_N)$. By our model we can describe interactions between these 
$I$-transformers only in the case in that
there is a possibility to choose the same $I$-time $t$ for all of them.
In this case we can consider the evolution of the system of the $I$-transformers
$\tau_1,\ldots,\tau_N$ as a trajectory in the $I$-space ${\bf Z}_p^N=
{\bf Z}_p\times\cdots\times
{\bf Z}_p,\; \; q(t) = (q_1(t), \ldots, q_N(t)).$

We think that this conditions of consistency for $I$-times of interacting
$I$-transformers plays the crucial role in many psychological experiments. We can not
obtain sensible observations for interactions between arbitrary individuals. There must be
a process of learning for the group $\tau_1,\ldots\tau_N$ which reduces $I$-times
$t_1,\ldots,t_N$ to the unique $I$-time $t$.

Thus, let us consider a group $\tau_1,\ldots,\tau_N$ of $I$-transformers with the
internal time $t$. The dynamics of $I$-states and motivations is determined by the
$I$-energy; $H(q,p),\ q\in {\bf Z}_p^N,\ p\in {\bf Z}_p^N$. It is natural to assume that
$$
H(q,p)=\sum_{j=1}^{N}\alpha_{j}p_{j}^2+V(q_1,\ldots,q_N),\ \alpha_j\in {\bf Z}_p.
$$
Here $H_{f}(p) = \sum_{j=1}^{N}\alpha_{j}p_{j}^2$ is the total energy of motivations for
the group $\tau_1,\ldots,\tau_N$ and $V(q)$ is the potential energy.
It is natural to choose $V(q)=\sum_{i\ne j}\Phi(q_i-q_j)$, where $\Phi(s),\ s\in {\bf Z}_p$, is
the potential of the interaction between $I$-transformers.

As usual, to find a trajectory in the phase $I$-space ${\bf Z}_{p}^N \times {\bf Z}_{p}^N$, we need to
solve the system of Hamiltonian equations:
\begin{equation}\label{hh}
	q_{j}=\frac{\partial H}{\partial p_{j}},\ p_{j}=-\frac{\partial H}{\partial q_{j}},
\; \;	q_{j}(t_0)=q_0,\ p_{j}(t_0)=p_0.
\end{equation}
(see [29] for such equations).

{\bf Consequences for cognitive and social sciences  and psychology}:

1. {\it Energy and information.} 
In our model a transmission of information is determined by the $I$-energy which is
the sum of  $I$-energy of motivations and potential $I$-energy. In principle, this
process  need no physical energy. Therefore, there might be transmissions of
information which could not be reduced to transmissions of physical energy. In this case we
cannot measure physical interactions 
(i.e., interactions in real space-time)
between two $I$-transformers, $\tau_1$ and $\tau_2$
(but we could measure an information interaction).
In particular, $\tau_1$ and $\tau_2$ can be individuals participating in psychological or social
experiments (or even experiments which exhibit anomalous behaviour).

2. {\it Distance and information.} $I$-processes may evaluate in an $I$-space
which differs from the real space (absolute Newton space or a space of general
relativity).
 Therefore the real ("physical") distance between $I$-transformers does not
play the crucial role in processes of  $I$-interactions.

3. {\it Time and information.}
Dynamics of information is dynamics with respect to $I$-time $t.$ There may be 
a correspondence $t_{\rm{phys}} = g(t)$ between real time $t_{\rm{phys}}\in {\bf R}$
and $I$-time $t\in {\bf Z}_p.$ This correspondence may not preserve distances. 
 
Let $\tau$ be an $I$-transformer having a continuous trajectory $q(t).$ Small variations
of $t, t^\prime= t+\delta t,$ imply small variations of $q:$
\begin{equation}
\label{V}
a^\prime= q(t^\prime) = a + p \delta t, \; a= q(t) .
\end{equation}

If (in some way) we find the internal time scale of $\tau,$ then it would be possible 
to find (via (\ref{V}) ) its $I$-state at the instant of time $t_{\rm{phys}}^\prime 
= g(t^\prime).$  If $t_{\rm{phys}}^\prime > t_{\rm{phys}}$ then such an $I$-measurement
can be considered as a prediction of future events; if  $t_{\rm{phys}}^\prime < t_{\rm{phys}}$ 
then we have  recalling. The relation (\ref{V}) gives only unsharp information. Thus such acts
of recalling and predictions may give a lot of unfruitful information.

4. {\it Motivation.} A motion in the $I$-space depends, not only on the initial $I$-state $q_0$, but also on
the initial motivation $p_0$. Moreover, the Hamiltonian structure of the equations of
motion implies that the motivation $p(t)$ plays the important role in the process of
the evolution. Thus $I$-dynamics is, in fact,  dynamics in phase $I$-space.

5. {\it Consistency for times.} An $I$-interaction between $I$-transformers is possible only 
if these $I$-transformers have consistent $I$-times.
Therefore every psychological or social
experiment has to contain an element of "learning" for $I$-transformers participating in
the experiment. A physical interaction need not be involved in such  learning. 
This can be any 
exchange of information between individuals (or a study of information about some
individual).

6. {\it Future and past.} The consistency condition for $I$-times 
does not imply such a condition for real
times, because different $I$-transformers can have different correspondence
laws for $I$-time and real time. For example, let us consider two
$I$-transformers, $\tau_1$ and $\tau_2$ satisfying the consistency condition for
$I$-times, i.e., $t_1=t_2=t$. We assume that it is possible to transform $I$-times of
$\tau_1$ and $\tau_2$ to real times $t_{1,\rm{phys}}=g_{1}(t_1)$ and 
$t_{2,\rm{phys}}=g_{2}(t_2)$. Let us also assume that $\tau_1$ and $\tau_2$ interact
by the $I$-potential $V(q_1-q_2)$, i.e., at the instant $t$ of $I$-time the potential
$I$-energy of this interaction equals $V(q_{1}(t)-q_{2}(t))$. If 
$t_{1,\rm{phys}}=g_{1}(t) \ne t_{2,\rm{phys}}=g_{2}(t)$ then such an
interaction is nothing than an interaction with the future or the past. 

7. {\it Social phenomena.} By our model any social group $G$ can be described by a system
$\tau_{1},\ldots,\tau_{N}$ of coupled $I$-transformers. There exists an $I$-potential
$V(q_1,\ldots,q_N)$ which determines an $I$-interaction between members of $G$.
For example, democratic societies are characterized by uniform $I$-potentials $V=\sum
\Phi(q_i-q_j)$. Here a contribution into the potential $I$-energy does not depend on an
individual. On the other hand, hierarchic societies are characterized by $I$-potentials of
the form:
$$ 
V=A_{0}\sum_{j \ne 0} \Phi(q_0,q_j)+ A_{1} \sum_{j\ne 0,1} \Phi(q_1,q_j)+ \cdots
$$
$$+
A_{k} \sum_{j \ne 0,\ldots,k} \Phi(q_k,q_j) + B \sum_{i,j\ne 0,\ldots,k} \Phi(q_i,q_j),
$$
where $|A_0|_p>>|A_1|_p>>\cdots>>|A_k|_p>>|B|_p$. These potentials describe the hierarchy 
$\tau_0\to\tau_1\to\cdots\to\tau_k\to(\tau_{k+1},\ldots,\tau_N)$. The $I$-transformer
$\tau_{0}$ can be a political, national or state leader or a God.

{\bf Remark 4.1.} (Transformers of information and classical real fields).
{\small  If $p\to \infty$ then 
the coding alphabet $\{0,1,...,p-1\}$ could be thought as being continuous, i.e., it
can be identified with the field of real numbers ${\bf R}.$ Therefore information space
$X= {\bf Z}_p, p\to \infty,$  can be identified with the infinite product of real fields,
$X= {\bf R}^\infty.$ Thus the $I$-state of an $I$-transformer $\tau$ can be identified
with a classical field $\phi(x), x \in {\bf R}$ (for example via Fourier coefficients).
Therefore  we can consider $I$-transformers as sources of
classical fields (in the limit $p\to \infty$). Of course, this is just a speculation, 
because we have no mathematical realization of this limiting procedure.}

\section{Information velocity, acceleration, mass and force, Newton's law.}

We have considered dynamics of  $I$-transformers of the unit mass. There the coefficient 
$v$ of a
proportion  between the variation $\delta q$ of the $I$-state and the variation $\delta
t$ of $I$-time $t$: $\delta q=v\delta t$, was
considered as a motivation. In the general case the motivation $p$ may not coincide with
$v$. Let us assume that the motivation $p$ is proportional to $v$, $p=mv, m\in {\bf Z}_p.$ This
coefficient $m$ of proportion  is called an $I$-{\it mass}. We also call $v$ an $I$-{\it velocity}.
Thus $\delta q=\frac{p}{m}\delta t$.

Let $\tau_1$ and $\tau_2$ be two $I$-transformers with the $I$-masses $m_1$ and $m_2$
and let $|m_1|_p>|m_2|_p$. Let $\tau_1$ and $\tau_2$ have the variations $\delta t_1$,
$\delta t_2$ of $I$-time of the same $p$-adic magnitude, $|\delta t_1|_p=|\delta t_2|_p$, and let these
variations generate the variations $\delta q_1$ and $\delta q_2$ of their $I$-states of the
same $p$-adic magnitude, $|\delta q_1|_p=|\delta q_2|_p$. To make such a change of the
$I$-state, $\tau_1$ need a larger motivation: $|p_1|_p=|\frac{\delta q}{\delta t}|_p
|m_1|_p>|p_2|_p=|\frac{\delta q}{\delta t}|_p |m_2|_p$. Thus the $I$-mass is a measure of
an {\it inertia of information.} We define a {\it kinetic} $I$-energy by $T=\frac{1}{2m}p^2$.

A variation $\delta t$ of $I$-time $t$ implies also a 
variation $\delta p$ of the motivation $p$:
$
	\delta p=f\delta t.
$
The coefficient $f$ of proportionality is called an $I$-{\it force.} Thus any change of the 
motivation is due to the action of an $I$-force $f$. If $f=0$ then $\delta p=0$ for any
variation $\delta t$ of $t$. Thus an $I$-transformer cannot change its motivation in
the absence of $I$-forces.

By analogue with the usual physics we call the coefficient $a$ of a proportion between the
variation $\delta v$ of the $I$-velocity $v$ and the variation $\delta t$ of the $I$-time
$t$, $\delta v=a \delta t$, an $I$-{\it acceleration.} Thus $\delta p = a m \delta t$. This
relation can be rewritten in the form of an information analogue of the second Newton law:
\begin{equation}\label{N}
	ma=f
\end{equation}
or
\begin{equation}\label{N1}
	\dot{p}=f.
\end{equation}
An $I$-force $f$  is said to be a {\it potential} force if there exists a
function $V(q)$ such that $f=-\frac{\partial V}{\partial q}$ where $V$ is called the
potential, or potential energy. The total $I$-energy $H$ is defined as the sum of the
kinetic and the potential $I$-energies, $H(q,p)=\frac{1}{2m}p^2 +V(q)$.
The Hamiltonian equation $\dot{p}=-\frac{\partial H}{\partial q}$ coincides with the
Newton equation $\dot{p} = f$.

{\bf Example 4.1.} (Hooke's $I$-system). Let the $I$-force $f$ be proportional to the 
$I$-state $q$, $f=m\beta^2 q$, where $m$ is the $I$-mass and $\beta \in {\bf Z}_p$
is a coefficient of the interaction. Here (\ref{N}) gives the equation $\ddot{q}=\beta^2 q$.
As $f= - \frac{\partial V}{\partial q}$, $V(q)=-\frac{m\beta^2}{2}q^2$ and
$H(q,p)=\frac{p^2}{2m}-\frac{m\beta^2 q^2}{2}$; the Hamiltonian equations are 
$\dot{q}=p/m$ and $\dot{p}=m\beta^2 q$. Their solutions have the form $g(t)=a e^{\beta
t}+b e^{-\beta t}$. By the condition (\ref{EXP}) the $I$-state $q(t)$ 
and motivation $p(t)$ are defined only for instants of $I$-time which satisfy the inequality 
\begin{equation}
\label{U}
	|\beta t|_p\leq r_p.
\end{equation}
This condition can be considered as a restriction for the magnitude of the 
$I$-force. If the coefficient of the interaction $|\beta |_p \leq r_p,$ then 
dynamics $q(t)$ of the $I$-state is well defined for all $t\in {\bf Z}_p.$
Larger forces imply the restriction condition for $I$-time. 
Let $|\beta|_p=1$. If $p\not=2$ then (\ref{U}) has the form $t\in U_{1/p}(0)$, i.e., 
$t=\alpha_1 p+\alpha_2 p^2+\cdots$. Thus the $I$-state $q(t)$ of the $I$-transformer
$\tau$ can be defined (observed) only for the instants of time 
$t_0 = 0, t_1 = p,..., t_{p-1} = (p-1) p,...\; .$ If $p=2$ then
(\ref{U}) has the form $t\in U_{1/4}(0)$, i.e.,
 and $t=\alpha_2 2^2+\alpha_3 2^3+\cdots$.
Thus the $I$-state $q(t)$ of 
$\tau$ can be defined (observed) only for the instants of time 
$t_0=0, t_1= 4, t_2= 8,...\; .$
 
Let $f=-m\beta^2 q$, i.e., $V(q)=\frac{m\beta^2 q^2}{2}$ and $\ddot{q}=-\beta^2 q$. Here
$q(t)$ and $p(t)$ have the form $g(t)=a \cos \beta t+b\sin \beta t$. Here we also have the
restriction relation (\ref{U}). As opposite  to the real case the $p$-adic trigonometric
functions are not periodical. There is no analogue of oscillations for the
$I$-process described by an analogue of Hooke's law. 

Let us consider the solution of the
Hamiltonian equations with the initial conditions $q(0)=0$ and $p(0)=m\beta$:
$q(t)=\sin\beta t$, $p(t)=m \beta\cos\beta t$. We have $qp=(m\beta/2)\sin 2\beta t$. By
using the $p$-adic equality $|\sin a|_p=|a|_p$ we get $|qp|_p=|m\beta|_p |\beta t|_p$. The
relation (\ref{U}) implies
\begin{equation}\label{U1}
	|q|_p |p|_p\leq |m\beta|_p r_p.
\end{equation}
This is a {\it restriction relation} for the trajectory $(q(t),p(t))$ in the phase $I$-space
(compare with [33]).
Let $\beta=1/m$. Then (\ref{U1}) gives $|q|_p |p|_p\leq r_p$. If the motivation $p$ is
strong $|p|_p=1$, then $q$ can be only of the form
$q=\alpha_1 p+\alpha_2 p^2+\cdots$, $p\ne 2$ and 
$q=\alpha_2 2^2+\alpha_3 2^3+\cdots$, $p=2$. If the motivation $p$ is rather weak then the
$I$-state $q$ of an $I$-transformer can be arbitrary.

The restriction relation (\ref{U1}) is natural if we apply our information model to
describe psychological (social) behaviour of individuals. Strong psychological
(social) motivations imply
some restrictions for possible psychological (social) states $q$. On the other hand, if 
motivations are rather weak an individual can, in principle, arrive to  any psychological
(social) state.

We discuss the role of the $I$-mass in the restriction relation (\ref{U1}). There the
decrease of the $I$-mass implies more rigid restrictions for the possible $I$-states (for
the fixed magnitude of the motivation). If we return to the psychological (social) 
applications we get
that the individual (or a group of individuals) with a small magnitude of $I$-mass and the
strong motivations will have quite restricted set of $I$-states.

The restriction relation (\ref{U1}) is an analogue of the Heisenberg uncertainty relations
in the ordinary quantum mechanics. However,  we consider a classical (i.e.,
not quantized) $I$-system. Therefore a classical $I$-system can have behaviour that is similar
to quantum behaviour.

\section{Mathematical "pathologies" in the formalism of the 
information mechanics and their interpretations}

In $p$-adic analysis the condition $f\equiv0$ does not imply that a differentiable
function $f$ is a constant, see [21], [46]. Therefore, there exist very complicated continuous motions
$(q(t),p(t))$ in the $I$-phase space for $I$-transformers with zero $I$-energy
($\dot{q}\equiv0$ or $\dot{p}\equiv0$). 

In psychological models these motions can be interpreted as motions without any motivation. Such
motions  need no information force. On the other hand, we can consider an
$I$-potential $V(q)$ such that $\frac{\partial V}{\partial q}=0$. Here the potential
$I$-energy $V(q)$ can have very complicated behaviour on the $I$-space $X={\bf Z}_p$.
At the same time the $I$-force $f=0$. Thus there may exist $I$-fields which do not induce any $I$-force.

All mathematical pathologies can be eliminated by the consideration of analytical
functions. If $f\equiv0$ and $f$ is analytic then $f=\textrm{constant}$.

In psychological models we can interpret analytical trajectories in the phase $I$-space 
as a "normal
behaviour", i.e., an individual need a motivation for the change of a psychological
state. Here we can observe some psychological (information) force which induces this
change. There is a psychological (information) field which generates this
force. 
The model puts trajectories (non-analytical) with zero motivation in relation with
abnormal psychological behaviour, mental diseases and
anomalous phenomena. Here an individual changes his psychological state without any
motivation in the absence of any information force. Here, in fact,
a $p$-adic generalization of the Hamiltonian formalism does not work. We need to propose
a new physical formalism to describe such phenomena.

Not all unusual properties of $p$-adic quantities are connected with non-analyticity.
For example, in $p$-adic analysis we can construct polynomials of the form
$V(x) = \alpha_0 + \alpha_1 x + \cdots+ \alpha_N x^N,$ where the coefficients
$\alpha_j$ are natural numbers, $\vert \alpha_j\vert_p=1$, such that $\epsilon= 
\sup_{x\in {\bf Z}_p} \vert V(x) \vert_p$  can be arbitrary small (see [46]). Therefore
the result of the simultaneous action of quite strong $I$-potentials $V_j(x)= \alpha_j x^j,
j=0,1,...,N,$ can have arbitrary small magnitude.

\section{Information work, conservation laws}

To eliminate from our consideration all "pathological" motions in the $I$-space, we shall
consider only $I$-quantities described by analytical functions. Of course, we do not claim
that only analytical functions describe real information processes. We like only to
simplify mathematical considerations.

Let $f(x)=\sum_{n=0}^{\infty}a_n x^n$, $a_n\in\mathbf{Q}_p$, and let the series converge
for $|x|_p\leq \delta$, $\delta=p^{\pm n},\ n=0,1,\ldots$. We define an integral of $f$ by
the formula (see [26]):
$$ \int_{a}^{b}f(x)\,dx=\sum_{n=0}^{\infty}\frac{a_n}{n+1}[a^{n+1}-b^{n+1}].$$
The series on the right-hand side converges for all $|a|_p,\ |b|_p\leq\frac{\delta}{p}$.
In particular, we can find an antiderivative $F$ of $f$ by the formula
$F(x)=\int_{0}^{x}f(x)\,dx$.

Let $f$ be an $I$-force which is described by the function $f(x)$ which is analytic for
$|x|_p\leq p$. Then this force is potential with the $I$-potential $V(x)=\int_0^x
f(x)\,dx$.

Let $\gamma=\{q(t),\ |t|_p\leq \lambda\}$ be an analytic curve in ${\bf Z}_p$. We
define its length element by $ds=v dt$, where $v=\dot{q}$ is the $I$-velocity. By
definition
$$W_{ab}=\int_{\gamma(a,b)}f\,ds=\int_{t_0}^{t_1}f(q(t))v(t)\,dt$$
where $q(t_0)=a$ and $q(t_1)=b$. The quantity $W_{ab}$ is said to be the {\it work} done by the
external $I$-force $f$ upon the $I$-transformer in going from the point $a$ to the point
$b$. By (\ref{N1}) we have
$$W_{ab}=\int_{t_0}^{t_1}m \dot{v}v\,dt=\frac{m}{2}\int_{t_0}^{t_1}\frac{d}{dt}v^2\,dt=
\frac{1}{2m}(p^2(b)-p^2(a)).$$
Thus the work done is equal to the change in the kinetic energy: $W_{ab}=T_b-T_a$. As the
$I$-force $f$ is potential then the work $W$ done around a closed orbit is zero:
$W=\oint f\,ds=0$. Thus the work $W_{ab}$ does not depend on an analytic trajectory
$\gamma(a,b)$.

We also have:
$$W_{ab}=\int_{\gamma(a,b)}-\frac{\partial V}{\partial
q}\,ds=\int_{\gamma(a,b)}-\frac{d}{dt}V(q(t) )\,dt=V(a)-V(b).$$
Thus $T_b-T_a=V(a)-V(b)$. We have obtained the energy conservation law for an
$I$-transformer: {\it if the $I$-forces acting on an $I$-transformer are described by 
analytical functions (in particular, they are potential), then the total energy of the
$I$-transformer, $H=T+V$, is conserved.}

At the moment the situation with nonanalytic potential $I$-forces is not clear. It may
be that the energy conservation law is violated in the general case. 

\section{Mechanics of a system of information transformers, constraints 
on information spaces}

Let $\tau_1,\ldots,\tau_N$ be a system of $I$-transformers with $I$-masses,
$m_1,\ldots,m_N\in{\bf Z}_p$. As in ordinary mechanics we must distinguish between the
external $I$-forces $F_i^{(e)}$ acting on $I$-transformers due to sources outside the
system and internal forces $F_{ji}$. As we have already discussed, $I$-times
$t_1,\ldots,t_N$ of $\tau_1,\ldots,\tau_N$ must satisfy the consistency condition:
\begin{equation}\label{C}
	t_1=t_2=\cdots=t_N=t.
\end{equation}
Thus the equation of motion for the $i$th particle is to be written:
\begin{equation}\label{S}
	\dot{p}_{i}=F^{(e)}_i+\sum_{j}F_{ji}.
\end{equation}
For some $I$-systems we may obey an information analogue of Newton's third law (a law of
information action and reaction): $F_{ij}=-F_{ji}$.

Set $x=\sum_i m_i x_i/M$, where $M=\sum m_i$. This point in the $I$-space is said to be
the {\it center of information} of the system. If the system satisfies Newton's third law for
$I$-forces then we get the equation of motion: $M\ddot{x}=\sum_i F^{(e)}_i=F^{(e)}$. The
center of information moves as if the total external $I$-force was acting on the
$I$-mass $M$ of the system concentrated at the center of information. We introduce the
motivation $P=M\dot{x}$ of the $I$-system. There is the following conservation theorem for
motions described by analytic functions $(q_j(t))_{j=1}^N,\ t\in{\bf Z}_p$: {\it if the 
total external $I$-force is zero, the total motivation of the $I$-system is conserved.}

{\bf Example 7.1.} (Social systems). We apply our $I$-model for describing a society 
$S$ which consists of individuals (or groups of individuals) $\tau_1,\ldots,\tau_N.$ 
There exist the center of information of $S$,
$x_S\in{\bf Z}_p$ which can be considered as a coding sequence for this society. If $S$
satisfies Newton's law of action-reaction for $I$-forces then its evolution is determined
by the external $I$-forces. If this evolution is not "pathological" then the motivation of
$S$ is conserved. Of course, there might be numerous "pathological" evolutions (for
example, evolutions with zero motivation, $P_S=0$).

For analytic motions the $I$-work done by all $I$-forces in moving the system from an
initial configuration $A=\{a_i=q_i(t_0)\}$ to a final configuration $B=\{b_i=q_i(t_1)\}$
is well defined:
$$W_{ab}=\sum_i \int_{\gamma(a_i,b_i)} F_i\,ds_i+\sum_{i\ne j}\int_{\gamma(a,b)}
F_{ji}\,ds_i$$
and $W_{ab}=T_B-T_A$, where $T=\frac{1}{2}\sum_i m_i v_i^2$ is the total kinetic $I$-energy of
the $I$-system. As usual $T=\frac{1}{2}M v^2+\frac{1}{2}\sum_i m_i v_i'^2$
where $v$ is the velocity of the center of information and $v_i'$ is the velocity of
$\tau_i$ with respect to the center of  information.

In our model of `social motion' (Example 7.1) we can say that the total kinetic 
energy of the society $S$
is the sum of the kinetic energy of the center of information of $S$ and the kinetic
energy of motions of individuals $\tau_j$ about the center of information.

We now consider the case when all $I$-forces are (analytical) potential:
$F_i^{(e)}=-\frac{\partial V_i}{\partial x_i}$ and $F_{ji}=-\frac{\partial
V_{ij}}{\partial x_i}$. To satisfy the law of action and reaction we can choose
$V_{ij}=\Phi_{ij}(x_i-x_j)$ where $\Phi_{ij}:{\bf Z}_p\to{\bf Z}_p, \Phi_{ij}= 
\Phi_{ji}$ are analytical
functions. Then by repeating the considerations of the standard mechanics over the reals we
obtain that $W_{AB}=-V(B)+V(A)$, where $V=\sum_i V_i+\frac{1}{2}\sum_{i,j}V_{ij}$ is the
total potential energy of the system of $I$-transformers. Therefore {\it the total $I$-energy
$H=T+V$ is conserved for every $I$-system with (analytical) potential $I$-forces (such
that $F_{ij}$ satisfy the law of information action-reaction).}

The consideration of $I$-systems induces dynamics in  multidimensional $I$-spaces;
$X_N={\bf Z}_p^N$. Such spaces can be useful for the description, not only systems of
$I$-transformers, but also individual $I$-transformers which have multidimensional
information spaces.

For example, let $\tau$ be a cognitive system and let $x=(x_1,\ldots,x_N),\ x_j\in{\bf Z}_p$, be a
set of ideas with which operates $\tau$
(i.e., there are $N$ parallel thinking processes $\pi_1,...,
\pi_N$ in $\tau$, see 
[18] - [20] for the details). Then the $I$-dynamics for $\tau$ is
described by the trajectory $(q(t),p(t))\in{\bf Z}_p^{2N}.$

As in standard mechanics, constraints play the important role in $I$-mechanics. The
simplest constraints ("holonomic") can be expressed as equations connecting $I$-states of
$I$-transformers $\tau_1,\ldots,\tau_N$ (or equations coupling different ideas in the
cognitive system):
$$f(q_1,\ldots,q_N,t)=0.$$
Here $f$ may be a function from ${\bf Z}_p^{N+1}$ into ${\bf Z}_p$ or a function
from ${\bf Z}_p^{N+1}$ into ${\bf R}$. The simplest constraints of the "real type"
are:\\
($C1$) $|q_1 - a|_p = r,\ldots, | q_N - a |_p = r,\ r>0, a \in {\bf Z}_p$, 
i.e., all $I$-transformers have to move over the
surface of the sphere $S_r(a)$;\\
($C2$) $|q_2-q_1|=r,\ldots,|q_N-q_1|=r$, i.e., there is the fixed $I$-transformer $\tau_1$
such that all other $I$-transformers must move on the distance $r$ from $\tau_1;$\\
($C3$) We can also consider an "information rigid body", i.e., a
system of $I$-transformers connected by constraints:
$
	|q_i-q_j|_p=r_{ij}.
$

{\bf Example 7.2.} (Restricted mentality). In cognitive sciences constraint ($C1$) can be used
for the description of a "restricted mentality". All ideas $q_1(t), ..., q_N(t)$ of 
a  cognitive system $\tau$ (generated by the
parallel processes $\pi_1,..., \pi_N$) belong to the restricted domain of ideas $X=S_r(a).$

{\bf Example 7.3.} (Ideology, religion). Let us consider the $I$-model of
a society $S$ with an ideology
(or religion) $a\in {\bf Z}_p.$ Then constraint ($C1$) can be interpreted as 
describing a social layer
${\cal L} = (\tau_1,..., \tau_N) $ of $S.$ These are all individuals who accept the
 ideology (or religion)
$a$ with an "information precision" $ r.$ Let this precision $r=1/p^k$ and
let $q_j(t) = (q_{j\alpha}(t))_{\alpha=0}^\infty, a=(a_\alpha)_{\alpha=0}^\infty.$ 
The constraint ($C1$) implies that 
$$
q_{j0}(t)= a_0,..., q_{j k-1}(t)= a_{k-1}, \; \mbox{but}\;
q_{jk}(t)\not = a_k.
$$
The members of ${\cal L}$ accept dogmas $a_0, ..., a_{k-1}$ of the ideology
(or religion), but they deny the dogma $a_k.$ In our hierarchical model
all other dogmas do not play any role. If the dogma $a_k$ is violated then the violation of
$a_{k+j}$ would not change a status of $\tau_{j}.$

{\bf Example 7.4.} (Evolution of an idea-fix). Let us consider a cognitive system
$\tau$ with $N$ parallel thinking processes $\pi_1,..., \pi_N.$ The constraint
($C2$) means that there is a thinking process in the cognitive system (in our case this is
$\pi_1$) which has a strong influence on all other thinking processes
$\pi_j, j \not=1.$ They could not go far away from $\pi_1.$ In psychology
$\pi_1$ may be interpreted as a process of evolution of an 
idea-fix. The constraint ($C2$)  in the $I$-space of the cognitive system $\tau$ implies
that all thinking activity of $\tau$ is connected with this idea-fix.

{\bf Example 7.5.} (Kingdoms, families and lovers).  The constraint ($C2$)
can be interpreted as describing a social layer ${\cal L} =(\tau_2,..., \tau_N)$
in a kingdom $K$ with the king $\tau_1.$ The evolution $q_1(t)$ of the
$I$-state of the king induces the information restrictions ($C2$) for evolutions of
$I$-states of members of the layer ${\cal L}.$ The same constraint may be used
for an information model of evolution of a family $F.$  Here $\tau_1$ may be
the father or mother. In the case $N=2$ we obtain the symmetric model
which  may be used for the description of a pair of lovers. Similar 
constraints in the $I$-space might explain some anomalous information connections between
individuals.

{\bf Example 7.6.} (Scandinavian society). The constraint ($C3$) may be used
in social sciences for the description of "Scandinavian societies". There
are nonzero distances $r_{ij} > 0, i,j= 1,..., N,$ between individuals
in the $I$-space. These distances are stable in the process of time evolution.

In the case of holonomic constraints described by the system of analytical functions:
$f_j:{\bf Z}_p^{N+1} \to {\bf Z}_p,\, j=1,\ldots, K$, i.e., $f_j(q_1,\ldots,q_N,t)=0$, we can
use the technique of the standard mechanics
\footnote{These methods may not be applied to constraints determined by real valued
functions. However, in the latter case we need not eliminate these constraints. These
constraints describe open subsets of the configuration $I$-space ${\bf Z}_p^N$. We can
choose such subsets as new configuration $I$-spaces.}.
If the equations are independent then we can introduce generalized $I$-coordinates
$\xi_1,\ldots,\xi_{N-K}$, and $q_l=q_l(\xi_1,\ldots,\xi_{N-K},t)$ $l=1,\ldots,N$, and 
$q_l(\xi,t)$ are analytical functions of $\xi$ and $t$
(see [21], [46] for the mathematical details).

{\bf Example 7.7.} (Hidden basic ideas).
If $q(t)=(q_l(t))_{l=1}^N$ 
describes ideas in the cognitive system at the instant $t$ of $I$-time, then by resolving constraints on 
these ideas we can find "independent ideas" $\xi(t)=(\xi_j(t))_{j=1}^{N-K}$ which, in fact, 
determine the $I$-state of the cognitive system. 

{\bf Example 7.8.} (Hidden leaders). If $q(t) = (q_l(t))_{l=1}^N$ describes the system 
$S=(\tau_1,\ldots,\tau_N)$ of $I$-transformers then the existence of generalized
$I$-coordinates $\xi$ can be interpreted as a possibility to reduce $I$-behaviour of $S$
to $I$-behaviour of the other system $G=(g_1,\ldots,g_{N-K})$ of $I$-transformers.

As in the standard mechanics, we introduce general $I$-forces:
\begin{equation}\label{G}
	Q_j=\sum_i F_i\frac{\partial q_i}{\partial \xi_j},
\end{equation}
where $F_i$ is the total $I$-force acting to $i$th $I$-transformer (i.e., $F_i=F_i^{(a)}+f_i$ 
is the sum of applied $I$-force $F_i^{(a)}$ and the $I$-force $f_i$ of constraints
\footnote{We can interpret $I$-constraints as unknown $I$-forces.}).

In our theory generalized $I$-forces have the natural interpretation (compare with the
situation with generalized forces in the usual mechanics). As we have noted, the
existence of generalized $I$-coordinates which are obtained from equations for constraints
means that the initial system $S=(\tau_1,\ldots,\tau_N)$ of $I$-transformers is
"controlled" by the other system $G=(g_1,\ldots,g_{N-K})$ of $I$-transformers. The
$I$-forces (\ref{G}) are, in fact, reaction $I$-forces, i.e., the control of $G$ over
$S$ generates $I$-forces applied to elements of $G$. By repeating of the usual
computations we get the equations of motion:
\begin{equation}
\label{GL}
\frac{d}{dt}(\frac{\partial T}{\partial \dot{\xi}_j})-\frac{\partial T}{\partial \xi_j}=Q_j,\,
j=1,\ldots,N-K.
\end{equation}
If the $I$-forces $F_i$ are potential with the analytical potential $V$, i.e., $F_i=
-\frac{\partial V}{\partial q_i}$, then generalized $I$-forces are also potential:
$Q_j=-\frac{\partial V}{\partial \xi_j}$. In this case the above equation can be written in
the form:
\begin{equation}
\label{L}
\frac{d}{dt}(\frac{\partial L}{\partial \dot{\xi}_j})-\frac{\partial L}{\partial \xi_j}=0,
\end{equation}
where $L=T-V$ is the $I$-Lagrangian.

It is important that the equations (\ref{L}) can be used to describe $I$-motions in the 
presence of an $I$-potential $V(q,\dot{q})$ which depends on (generalized) $I$-velocities
$v_i=\dot{q}_i$. In this case
\begin{equation}
\label{GF}
Q_j=-\frac{\partial V}{\partial q_j}+\frac{d}{dt}(\frac{\partial V}{\partial \dot{q}_j})
\end{equation}
and $L=T-V$.

These velocity-dependent potentials may play an important role in $I$-processes.
In particular, there might have applications in such an exotic field as 
anomalous phenomena.
It is claimed (see, for example, [12] - [17]) that a psychokinesis effect can be observed
for some random physical processes and it cannot be observed for deterministic processes.
It might be tempting to 
explain this phenomenon on the basis of the assumption that an $I$-field generated
in experiments on the psychokinesis corresponds to a potential which depends on the
$I$-velocity (thus the corresponding $I$-force is defined by (\ref{GF}) ). The $I$-velocity
is higher for random processes. Therefore the interaction is stronger for these processes.

\section{Quantum mechanics on information spaces}

It is quite natural to quantize classical mechanics on information spaces over 
${\bf Z}_p$. We give the following reasons for such quantization. 
Observations over $I$-quantities are statistical observations.
We have to study statistical ensembles of $I$-transformers
(instead studying of an individual $I$-transformer). Such statistical ensembles are
described by quantum states $\phi$. As usual in quantum formalism, we can assume that a
value $\lambda$ of an $I$-quantity $A$ can be measured in the state $\phi$ with some
probability ${\bf P}_{\phi}(A=\lambda)$. This ideology is nothing than the application of the
{\it statistical (ensemble) interpretation of quantum mechanics} (see, for example,
[44] or [6]) to the information theory. By this
interpretation any measurement process has two steps:
(1) a preparation procedure ${\cal E}$; (2) a measurement of a quantity $B$ in the
states $\phi$ which were prepared with the aid of ${\cal E}$. 

Let us consider these steps in the information framework. By ${\cal E}$ we have to
select a statistical ensemble $\phi$ of $I$-transformers on the basis of some
$I$-characteristics. Typically in quantum physics a preparation procedure ${\cal E}$ is
realized as a filter based on some physical quantity $A$, i.e., we select elements which
satisfy the condition $A=\mu$ where $\mu$ is one of the values of $A$. We
can do the same in quantum $I$-theory. An $I$-quantity $A$ is chosen as a filter, i.e.,
$I$-transformers for the statistical ensemble $\phi$ are selected by the condition
$A=\mu$ where $\mu\in {\bf Z}_p$ is some information. For example, we
can choose $A=p$, the motivation, and select a statistical ensemble 
$\phi=\phi(p=\mu)$ of $I$-transformers which have the same motivation
$\mu\in{\bf Z}_p$. Then we realize the second step of a measurement process and
measure some information quantity $B$ in the state $\phi_{(p=\mu)}$. For example, we
can measure the $I$-state $q$ of $I$-transformers belonging to the statistical ensemble
described by $\phi_{(p=\mu)}$. We shall obtain a probability distribution 
${\bf P}(q=\lambda|p=\mu),\ \lambda,\mu\in {\bf Z}_p$ (a probability that $I$-transformer
has the $I$-state $q=\lambda$ under the condition that it has the motivation $p=\mu$).
It is also possible to measure the $I$-energy $E$ of $I$-transformers. We shall obtain a
probability distribution ${\bf P}(E=\lambda|p=\mu),\ \lambda,\mu\in{\bf Z}_p$\footnote{
We now try to provide theoretical foundations for quantum $I$-theory.
We do not discuss  concrete measurement procedures for $I$-quantities.
In particular, at the moment it is not clear how the $I$-energy can be measured. It seems
natural to use an analogue with usual quantum theory here. The $I$-energy can be measured
in the process of interactions between $I$-transformers or interactions of
$I$-transformers and $I$-fields.}.
On the other hand, we can prepare a statistical ensemble $\phi_{(q=\mu)}$ by fixing some
information $\mu\in{\bf Z}_p$ and selecting all $I$-transformers which have the
$I$-state $q=\mu$.  Then we can measure motivations of  these $I$-transformers and we
shall obtain a probability distribution ${\bf P}(p=\lambda|q=\mu)$.

Other possibility is to use a generalization of the 
individual interpretation of quantum mechanics. By this interpretation a wave function $\psi(x), x \in {\bf R}^n,$
describes the state of an individual quantum particle. In the same way we may assume
that a wave function $\psi(x), x\in {\bf Z}_p^n,$ on the $I$-space describes the state of
an individual $I$-transformer $\tau.$ 

{\bf Example 8.1.} (Referendum). In some social models we can consider individuals as quantum
$I$-transformers. A referendum is one of the possible measurement devices. 
Here the act of a measurement is a procedure of giving answers to questions of 
the referendum. By the individual interpretation individuals have no
definite answers to these questions before the referendum. These answers 
(information communications) are created in the process of the referendum. In fact, this
$I$-measurement changes $I$-states of individuals.

{\bf Example 8.2.} (Conscious measurement of quantum subconsciousness) We might describe
brain's functioning by the following quantum $I$-model. There is a quantum system,
subconsciousness\footnote{By our model [18] the subconsciousness is a kind of processor
in that work a large number of dynamical systems of the form $x_n= f(x_{n-1}),$
where $f: {\bf Z}_p \to {\bf Z}_p$ is a continuous function. Attractors of these
dynamical systems are solutions of problems.}, which state is described by the wave function
$\psi(x), x \in {\bf Z}_p.$ There is a measurement device, consciousness, which measures the
$I$-state $q$ of the subconsciousness. The concrete value (idea) of $q$  is not determined before the act 
of the conscious measurement. It is created only at the instant of a measurement. 
Of course, this act of a measurement (as in the ordinary quantum mechanics) changes
the state of the subconsciousness. The main difference from the standard quantum mechanical
scheme is that
we consider repeatable measurements over the same quantum system. In ordinary 
scheme of a quantum measurement we consider an ensemble of identical systems. 
At the moment we can present only some speculations about nature of the consciousness.
The consciousness is an information field generated by the brain. This field interacts 
continuously with the subconsciousness\footnote{This is a feedback process [18]:
the conscious field sends to the subconsciousness a problem $x_0$ which is the
initial condition for one of dynamical systems located in the subconsciousness 
(this is the signal to start the work of the dynamical system). On the other hand,
an attractor of this dynamical system (a solution of the problem $x_0$)
 interacts with the conscious field (this is 
the signal to stop the work of the dynamical system).}. 

{\bf Example 8.3.} ( Psychoanalysis). On the basis of the model of the previous example
we can interpret psychoanalysis as a series of measurements of the $I$-state of the
subconsciousness. These measurements continuously change the wave function of the
subconsciousness. Thus psychoanalysis is a treatment based on the series of quantum 
$I$-measurements. In fact, psychoanalytic tries to provide some functions of
the conscious field.\footnote{Thus Freud's theory [47] may be interpreted in the following way.
If the interaction between the consciousness and subconsciousness is not sufficiently strong,
the consciousness "cannot see" some attractors of dynamical systems located 
in the subconsciousness.
Thus the consciousness cannot send to the subconsciousness
the signal to stop iterations of these dynamical systems. These dynamical systems
are continuously busy and they cannot be used for other purposes. Other possibility
is that the general interaction is strong. However, the consciousness "cannot recognize"
some attractor as a solution of a problem because a strong external information field
(a taboo) might hinder to the interaction. Therefore a psychoanalytic has to 
find the hidden attractor and by  this
act the work of the corresponding dynamical system will be stoped. Of course, he must
be isolated from the corresponding "taboo-field".}.

The problem of interpretations is an important problem of ordinary quantum 
mechanics on real space. The same problem arises 
immediately in our quantum $I$-theory. We do not like
to start our investigation with a hard discussion on the right interpretation. 
We can be quite
pragmatic and use both interpretations by our convenience. However, the reader, who is interested
in foundations of quantum mechanics, can find the extended discussion on the problem of
the interpretation in Appendix 2.

In fact, a mathematical model for quantum $I$-formalism has  been already constructed. This
is  quantum mechanics with $p$-adic valued functions, see [29], [18], [37], [34], [35]. 
We present briefly this model. The space of quantum states is
realized as a $p$-adic Hilbert space ${\cal K}$ (see [29], [18] about the theory of such
spaces). This is a $\bf{Q}_p$-linear space which is a Banach space (with the norm
$\|\cdot\|$) and on which is defined a symmetric bilinear form
$(\cdot,\cdot):{\cal K}\times{\cal K}\to \bf{Q}_p$. This form is called an inner
product on ${\cal K}$. It is assumed that the norm and the inner product are connected
by the Cauchy-Bunaykovski-Schwarz inequality:
$|(x,y)|_p\leq \|x\| \|y\|,\ x,y\in{\cal K}.$

{\bf Remark 8.1} It is possible to use more general spaces over different extensions of
$\bf{Q}_p$ (analogues of complex Hilbert spaces). 

By definition quantum $I$-state
$\phi$ is an element of ${\cal K}$ such that $(\phi,\phi)=1$; quantum $I$-quantity $A$
is a symmetric bounded operator $A:{\cal K}\to{\cal K}$, i.e., $(Ax,y)=(x,Ay),\
x,y\in{\cal K}$\footnote{In $p$-adic models we do not need to consider unbounded
operators, because all quantum quantities can be realized by bounded operators, 
see [29], [18], [37], [34], [35].}.
We discuss a statistical interpretation of quantum states in the case of a discrete
spectrum of $A$.

Let $\{\lambda_1,\ldots,\lambda_n,\ldots\},\ \lambda_j\in{\bf Z}_p$ be eigenvalues of
$A$, $A\phi_n=\lambda_n\phi_n,\ \phi_n \in {\cal K},\ (\phi_n,\phi_n)=1$. The eigenstates
$\phi_n$ of $A$ are considered as pure quantum $I$-states for $A$, i.e., if the system of
$I$-transformers is described by the state $\phi_n$ then the $I$-quantity $A$ has the
value $\lambda_n\in {\bf Z}_p$ with probability 1. Let us consider a mixed state
\begin{equation}\label{QS}
	\phi=\sum_{n=1}^{\infty}q_n \phi_n,\ q_n \in {\bf Q}_p,
\end{equation}
where $(\phi,\phi)=\sum_{n=1}^{\infty}q_{n}^2=1$\footnote{As in the usual theory of
Hilbert spaces, eigenvectors corresponding to different eigenvalues of a symmetric
operator are orthogonal.}. By the statistical interpretation of $\phi$ if we realize a
measurement of the $I$-quantity $A$ for $I$-transformers belonging to the statistical
ensemble described by $\phi$ then we obtain the value $\lambda_n$ with probability
$P(A=\lambda_n|\phi)=q_{n}^{2}$.

The main problem (or the advantage?) of this quantum model is that these probabilities
belong to the field of $p$-adic numbers ${\bf Q}_p$. The simplest way is to eliminate 
this problem by considering only finite mixtures (\ref{QS}) for which $q_n \in 
{\bf Q}_p$ (the
field of rational numbers $\bf{Q}$ is a subfield of ${\bf Q}_p$). In this case the
quantities ${\bf P}(A=\lambda_n|\phi)=q_{n}^2$ can be interpreted as usual probabilities (for
example, in the framework of Kolmogorov's theory [48]). Therefore we may assume that there
exist (can be prepared) quantum $I$-states $\phi$ which have the standard statistical
interpretation: when the number $N$ of experiments tends to infinity, the frequency
$\nu_{N}(A=\lambda_n|\phi)$ of an observation of the information
$\lambda_n\in{\bf Z}_p$ tends to the probability $q_{n}^2$.

However, we can use a more general viewpoint to this problem. In book [29] a (non-Kolmogorov)
probability model with $p$-adic probabilities has been developed. 
If we use a $p$-adic generalization of a frequency approach to probability
(see R. von Mises, [49]), then $p$-adic probabilities
are defined as limits of relative frequencies $\nu_{N}$ with respect to the
$p$-adic topology \footnote{It is quite surprising that in the $p$-adic framework we can obtain
negative rational frequency probabilities
[29], [50]. On the other hand, negative `probabilities'
appear in quite natural way in many quantum models (see, for example, [51]- [53]).
P.A.M. Dirac was the first to introduce explicitly the concept of negative probability
(in close connection with the concept of negative energy), [51]. 
R. Feynman also discussed the possibility to use negative probabilities
in quantum formalism, see [52]. In particular, he
remarked: "The only difference between a
probabilistic classical 
world and the equations of the quantum world is that somehow or other it
appears as if the 
probabilities would have to go negative, and that we do not know, as far as
I know, how to 
simulate".   These probabilities were used to explain violations of Bell's
inequality (see review [53]). Here the assumption
that a distribution of hidden variables may be a signed `probabilistic
measure' implies existence of numerous models with hidden variables
in that Bell's inequality is violated. Wiegner's distribution on the phase
space gives other example of signed quantum `probabilistic' distribution.
In works [38] - [40], [54], [55]
the $p$-adic probabilities (which are well
defined on the mathematical level of rigorousness)
were used to justify the use of negative probabilities in quantum theories.}. 
The relative frequencies $\nu_{N}\in\bf{Q}$ and they can be
considered, not only as elements of $\bf{R}$, $\bf{R}\subset\bf{Q}$, but
also as elements of ${\bf Q}_p$, ${\bf Q}_p\subset\bf{Q}$. 

By using the $p$-adic frequency probability model for the statistical interpretation of
quantum $I$-states we may assume that there exists $I$-states $\phi$ (ensembles of
$I$-transformers) such that the relative frequencies $\nu_{N}(A=\lambda_n|\phi)$ have no
limit in $\bf{R}$, i.e., we cannot apply the standard law of the large numbers in this
situation. Hence if we realize measurements of an $I$-quantity $A$ for such
a quantum $I$-state and study the observed data by using the standard statistical
methods (based on real analysis),
then we shall not obtain the definite result. There will be
only random fluctuations of relative frequencies, see [29], [50] .

{\bf Remark 8.2.} Such a behaviour can be related to  psychological  
experiments. Here the possibility of the use of $p$-adic probability
models gives the important
consequence for scientists doing experiments with a statistical $I$-data:
{\it the absence of the statistical stabilization (random fluctuation) does not imply the
absence of an $I$-phenomenon. This statistical behaviour may have the meaning that this
$I$-phenomenon cannot be described by the standard Kolmogorov probability model}.

We now discuss other interesting implications of $p$-adic probability theory. There exists
statistical samples [29], [50] in which the frequencies $\nu_{N}\to 0$, in the standard real
topology, but $\nu_{N}\to\alpha \not= 0$ in ${\bf Q}_p.$  In this case the
usual (Mises) frequency probability ${\bf P} (A=\lambda|\phi)=0$. This implies that we have to
consider the event $\{A=\lambda|\phi\}$ (an observation of the information $\lambda$) as
nonphysical event. However, from the point of view of the $p$-adic probability theory this
is the physical event (of course, in the sense of $I$-physics). 

The evolution of a $p$-adic wave function is described by an $I$-analogue of the
Schr\"odinger equation:
\begin{equation}
\label{Sch}
\frac{h_p}{i}\; \frac{\partial \psi}{\partial t} (t,x) =
\frac{h_p^2}{2m}\; \frac{\partial^2 \psi}{\partial x^2} (t,x) -
V(t,x) \psi(t,x),
\end{equation}
where $m$ is the $I$-mass of a quantum $I$-transformer. Here a constant
$h_p$ plays the role of the Planck constant. By pure mathematical reasons 
(related to convergence of $p$-adic exponential and trigonometric series)
it is convenient to choose $h_p= \frac{1}{p}.$ 

We may also present some physical arguments for such a choice.
In ordinary quantum mechanics the Planck constant is related to the measure of discretization.
The constant $h_p= \frac{1}{p}$ is related to the level of discretization of information.

 If we use the statistical interpretation of 
quantum mechanics then the parameter $t$ plays the role of common $I$-time
for elements of a statistical ensemble of $I$-transformers described by the wave function.
Therefore, to be able to describe the evolution of a quantum state $\psi,$
we must have consistent $I$-times for elements of this statistical ensemble.

We use the factor $i=\sqrt{-1}$ in (\ref{Sch}), because we like to have the
total coincidence with formulas of the ordinary quantum mechanics.
As we have already noted, in the $p$-adic case
the functions
$e^{i \alpha x}$ and $e^{\alpha x}$ have the same (non-oscillating) behaviour.
Therefore, in principle, we can use the analogue of (\ref{Sch}) in that
the factor $i$ is omitted. 

The use of $i$ implies the consideration of the extension 
${\bf Q}_p(i)={\bf Q}_p\times i {\bf Q}_p$ of ${\bf Q}_p.$ Elements of this
extension have the form $z=a+ i b, a,b\in {\bf Q}_p.$ This extension is well defined for
$p= 3, {\rm{mod}}\; 4.$ As usual, we introduce a congugation $\overline{z}=
a- i b;$ here we have $z\overline{z}= a^2 + b^2.$ In what follows we assume that 
wave functions take values in ${\bf Z}_p(i)={\bf Z}_p\times i {\bf Z}_p.$

{\bf Example 8.4.} ( A free $I$-transformer). Let the potential $V=0.$ Then 
the solution of the Schr\"odinger equation corresponding to the $I$-energy
$E= \frac{{\bf p}^2}{ 2 m}$ has the form\footnote{We note that formal expressions
for analytical solutions of $p$-adic differential equations coincide with the corresponding 
expressions in the real case (in fact, we can consider these equations over
arbitrary number field, see [29]). However, behaviours of these solutions 
are different.} :
\begin{equation}
\label{W}
\psi_{{\bf p}}(t,x)= e^{i({\bf p} x - E t)/h_p}.
\end{equation}
By the choice $h_p=1/p$ this function is well defined for all $x\in {\bf Z}_p$
and $t \in {\bf Z}_p$. As $\psi\overline{\psi} \equiv 1,$
this wave function describes the uniform ($p$-adic probability) distribution,
see [29], on the ring of $p$-adic integers ${\bf Z}_p.$ Thus an $I$-transformer 
$\tau$ in the state $\psi$ can be observed with equal probability in any state
$x\in {\bf Z}_p.$ In this sense behaviour of a free $I$-transformer is similar 
to behaviour of the ordinary free quantum particle. On the other hand, there is no 
analogue of oscillations:
$ 
\psi_{{\bf p}}(t,x)= \cos ({\bf p} x - E t)/h_p + i \sin ({\bf p} x - E t)/h_p,$
and $\vert \cos ({\bf p} x - E t)/h_p\vert_p=1, \vert\sin ({\bf p} x - E t)/h_p\vert_p=
\vert  ({\bf p} x - E t)/h_p\vert_p.$

{\bf Remark 8.4.} Is it possible to reproduce oscillations 
with respect to  ordinary real time on the basis
of the information model? It could be done by a time scaling. Let 
$f:{\bf Z}_p \to {\bf Z}_p$ be an arbitrary continuous function. Then
$f(t+ k p^n)\approx f(t)$ for all $k\in {\bf Z}$ for sufficiently
large $n$ (uniformly for $t\in {\bf Z}_p$). Let $t_{{\rm{phys}}}= g(t)$
 be a law of the correspondence between $I$-time $t \in {\bf Z}_p$ and real time
$t_{{\rm{phys}}}\in {\bf R}.$ If $2\pi= g(p^n)$ then the $p$-adic continuity will 
imply the periodicity in real time. Therefore, 
the ordinary wave behaviour is nothing other than a consequence of continuity of 
information flows and the appropriative choice of a time scale. Depending 
on a time scale an $I$-process may  or may not exhibit wave behaviour in the real
picture of reality.

We consider a psychological (and social) consequence of Example 8.4:
{\it in the absence of the external potential the same motivation ${\bf p}$
may imply any $I$-state $x\in {\bf Z}_p.$}

Let us consider mixtures of states of the form (\ref{W}). We set $t=0.$
Let $\psi(x)= a_1 \psi_{{\bf p}_1} +  a_2 \psi_{{\bf p}_2},\; a_1, a_2\in {\bf Z}_p.$

If we compute $< \psi,\psi>=\int_{{\bf Z}_p} \psi(x) \overline{\psi(x)} d x$
(where $d x$ is a uniform $p$-adic valued distribution on
${\bf Z}_p$) we see a large difference with ordinary quantum mechanics: 
$< \psi,\psi> \not= a_1\bar{a}_1 + a_2\bar{a}_2 .$ There is nonzero correlation term. 
For $\alpha= ({\bf p}_1 - {\bf p}_2)/ h_p,$ we have [18]:
$$
T(\alpha) = < \psi_{{\bf p}_1}, \psi_{{\bf p}_2}> +< \psi_{{\bf p}_2}, \psi_{{\bf p}_1}>=
\frac{\alpha \sin\alpha}{1- \cos \alpha}.
$$
Thus there are correlations between the motivations ${\bf p}_1$ and 
${\bf p}_2$ in the state $\psi.$ By using the individual interpretation of
quantum mechanics we say that an $I$-transformer $\tau$ with the wave function $\psi$
is in the superposition of two motivations ${\bf p}_1$ and ${\bf p}_2.$ Moreover,
these motivations could not be measured exactly (compare with [18], [33]). 

Such a situation is natural for psychological and social phenomena. In fact, a psychological
or social motivation may be not represented in the brain in the definite form before the act of
a measurement (at least for some quantum information states). Moreover, 
it cannot be measured exactly.
Such information measurements may be used as illustrations of the process of a measurement 
in ordinary quantum mechanics. By analogy we can say that the definite value of a physical 
observable is created in a long process of the interaction with an equipment. Moreover,
it can be never measured exactly (compare with [56] - [57]).

{\bf Example 8.5.} (Quantum Hooke's system) To give an example of a Hamiltonian with 
discrete spectrum, we consider the formal $p$-adic generalization of the Hamiltonian
of a harmonic oscillator:
$$
\hat{H} = - \frac{h_p^2}{2 m} \frac{d^2}{d x^2} - \frac{1}{2} m \omega^2 x ^2 -\frac{1}{2},
$$
where $m$ is the $I$-mass. We consider $\omega$ simply as the coefficient of interaction
(there is no analogue of harmonic oscillations). The operator $\hat{H}$ has eigenvalues
$E_n = h_p \omega  n , n=0,1,...$ (see [18]). 
However, in the $p$-adic case the difference between continuous and discrete
spectra is not so strong (for each $ E_n ,$  
we have  $E_n= \lim_{k\to\infty} E_{l_k}, l_k\not= n$). On the other hand, discreetness of a
spectrum, of course, induces some restrictions on  values (information) which can be observed.

\section{Appendix 1: Models of reality and number systems}

{\small Since Newton's time, we use a model of physical reality based on a description of all
physical processes by real numbers. In fact, the use of real numbers is equivalent to the
assumption that any physical quantity can be measured (at least in principle) with an
infinite precision. We shall discuss this point more carefully.

To realize a measurement of a physical quantity $x$, first we have to fix a unit of a
measurement $l=1$. We assume that there exists such a natural number $n$ that
\begin{equation}\label{A}
	(n-1)l\leq x<nl.
\end{equation}
This assumption is a mathematical postulate, the {\it Archimedean axiom.} Therefore by (\ref{A})
we restrict our considerations to physical phenomena which can be described on the basis
of the Archimedean mathematical model.

We now consider the next step of the measurement process. If $y_1=(n-1)l\ne x$ then we have to
measure the quantity $x_1=x-y_1$ by using a smaller unit of the measurement. Typically we
fix a natural number $m>1$ ( the scale of the measurement) and choose the new unit
$l_1=l/m$. Then we apply the Archimedean axiom (\ref{A}) to the quantities $x_1$ and $l_1$
and obtain a natural number $\beta_1$ $(\beta_1=1,\ldots,m))$: $(\beta_1-1)l_1\leq
x_1<\beta_1 l_1$. This procedure can be continued. If $y_2= ( \beta_1- 1 ) l_1 \ne x_1$ then we can
use the new unit of measurement $l_2=l_1/m$ to measure the quantity
$x_2 = x_1 - y_2 $ and so on. We remark that
\begin{equation}\label{P}
	x=(n-1)+x_1=(n-1)+\frac{\alpha_{-1}}{m}+x_2=n-1+\frac{\alpha_{-1}}{m}+\cdots+
	\frac{\alpha_{-n}}{m^n}+x_{n+1},
\end{equation}
where $\alpha_{-k}=\beta_k-1=0,1,\ldots,m-1$.
To obtain the real numbers model for physical reality, we assume that the above process of
measurements of every physical quantity $x$ can be continued by an infinite number of
steps. We call this postulate the postulate of an {\it infinite precision} of measurements or
the {\it Newton axiom.} By this axiom any physical quantity $x$ can be identified with the real
number:
\begin{equation}\label{R}
	x=\cdots+\frac{\alpha_{-n}}{m^n}+\cdots+\frac{\alpha_{-1}}{m}+ \alpha_0+\alpha_1
	m+\cdots+\alpha_k m^k=\alpha_k\cdots \alpha_0,\alpha_{-1}\cdots\alpha_{-n}\cdots,	 
\end{equation}
where $\alpha_{\pm j}=0,1,\ldots,m-1$,
(here the number $(n-1)$, see (\ref{P}), is also expanded with respect to powers of $m$).

Both the Archimedean and Newton axioms are natural for the description of an extended class
of physical phenomena. 
The basis of the Archimedean-Newton model of
reality is Newton's space which is continuous, infinitely divisible and
infinitely deep. All physical objects are located in this space and their location can be
determined (at least in principle) with an infinite precision. 

It would be natural to develop other models of physical reality which are not based on the 
Archimedean and Newton axioms. 

The quantum formalism is one of  successful attempts to give a new model of physical
reality. The Archimedean and Newton axioms cannot be applied to quantum observables.
However, quantum theory uses the old mathematical basis, real numbers. Of course, such a
situation when non-Archimedean and non-Newtonean phenomena are considered in "real"
reality should induce paradoxes. One of such paradoxes is the EPR paradox which gives the
right consequence, the {\it death of reality } (see, 
for example, [6],[7] for the details). 

In the present paper we started to develop a new model of reality, information reality, based
on systems of $m$-adic numbers. This is a {\it non-Archimedean model.} Here we
could not `compare' two arbitrary information quantities $x, y \in {\bf Z}_m.$ This is
a {\it non-Newtonian model.} Information could not be `measured with an infinite
precision.' Information spaces are discrete. There always exists a `minimal space
length', a bit of information\footnote{Our model of reality is closely connected with physical
theories based on the fundamental length formalism or discrete space-time, see, 
for example [59]-[61]. }

\section{Appendix 2. An interpretation of 
quantum information mechanics.}

\small We shall discuss the problem of interpretation on the basis of an example
of a cognitive quantum system.  Our analysis of measurement processes
for quantum cognitive systems implies that we have to use the ensemble 
interpretation. There are two main viewpoints on the ensemble interpretation.
The first is based on {\it realism} (see, for example, [44]). Here every physical quantity $A$ 
pertaining to a quantum state $\phi$ which 
describes a statistical ensemble $S$ has some definite value for each
$s\in S.$ The second is based on {\it empiricism.} Here it is not assumed 
the "objective existence" of definite values of $A.$ The $\phi$
gives only probabilities that $A$ would take some values (if a measurement
of $A$ is performed). In fact, in conventional quantum mechanics
the ensemble interpretation is based on the stronger form of empiricism:
it could not be assumed that $A$ has definite "objective value" $a$ for each
$s \in S.$ Our analysis implies rather strange consequences.
On the one hand, we understood that 
it is impossible to interpret a quantum state $\phi$ on the basis of
pure realism. On the other hand, we could not follow the conventional ensemble interpretation.
On the one hand, a preparation procedure ${\cal E}_b$ which can be realized as a filter 
with respect to values $\{ b_j\}$ of an $I$-quantity $B$ produces a statistical ensemble
$S$ (described by the quantum state $\phi$) in which each individual $I$-system
has some fixed (objectively existing) value $b_j.$ On the other hand, there exist
$I$-quantities for which we cannot assume that their have definite ("objective") values for
elements $s \in S.$ If $A$ is such an $I$-quantity, then its values are generated in the process 
of a measurement ${\cal M}_a$ over elements of the $S.$ This measurement does not
imply a discontinuous collapse of the state $\phi$ to the state $\phi_a$ corresponding to 
the fixed value $a$
of $A.$ In the contrary this value $a$ of $A$ is created in the long process of the 
interaction between
an $I$-system and a measurement device\footnote{Thus the $\phi$ is also related to individual
$I$-systems.}. There is the clear evidence that at least cognitive systems have hidden
$I$-variables
($I$-states of the brain which are represented by configurations of
excited neurons, see, for example, [62], [63]) which exist objectively and 
determine with some probabilities results of the ${\cal M}_a.$ In fact, these 
hidden $I$-variables (at least some of them) are not longer hidden. The modern experimental
neuroscience gives the possibility to observe configurations of excited neurons 
corresponding to different reactions (results of ${\cal M}_a),$
see, for example, [64], [65] on experiments (based on functional magnetic resonance imaging
machine) for memory neurons configurations.

{\bf Remark 10.1.} (Distribution of cognitive information in real space)
 Let $k \geq m >1$ be natural numbers. We introduce a map
$$
j_{mk} : {\bf Z}_m \to [0,1], x= \sum_{l=0}^\infty \alpha_j m^j \to 
x_{{\bf R}}= \sum_{l=0}^\infty \frac{\alpha_l}{k^{l+1}}.
$$
We can present the speculation that one of  maps $j_{mk}$ 
gives the spatial distribution of information in a cognitive system
(in the case of ``one dimensional brain").  This spatial distribution
can have quite exotic structure. For example,
the image $j_{mk}(U_r(a))$ of a ball $U_r(a)$ can be a {\it fractal}
(a kind of dusty set) in the real space.  This model can have some connection
with {\it frequency domain methods} [66] in that populations of cortical oscillators 
self-organize by frequencies; same-frequency sub-populations of oscillators can interact 
in the sense that a change in phase deviation in one will be felt by the others in the 
sub-population.
Thus here the spatial nearness of neurons (and even the existence of synaptic 
connections between two neurons) does not guarantee that they interact.

To simplify our considerations, we consider only states which provide the conventional
(Kolmogorov) probability interpretation.

In what follows students can be considered 
as analogues of quantum particles and professors as analogues of measurement
devices. Students of a University have to pass a test ${\cal M}.$ They have to give 
n answer to the question
$L.$ Denote the set of all possible answers $\{a_j\}$ to $L$ by the symbol
${\cal A}.$ If we use the coding alphabet $\{0,1,..., p-1\},$ then the elements of 
${\cal A}$ can be presented by $p$-adic integers. To prepare to this test, students
have to read one of books  ${\cal B} =\{ b_j\}$ (here $b_j \in {\bf Z}_p$ are coding sequences
for books); a student has no time for reading of more than one book.
 All books give a description of the subject; but these descriptions are not
identical.  The process of reading is considered as a preparation procedure ${\cal E}_b.$
It produces a statistical ensemble $S$ of students who have read a book $b_j \in 
{\cal B}.$ The ${\cal E}_b$ can be considered as a filter on the set of all students of University.
By quantum formalism (with the ensemble interpretation)
$S$ is described by a quantum state $\phi$
(which is a vector in a $p$-adic Hilbert space). This state is presented in the form:
\begin{equation}
\label{xe}
\phi = \sum c_j \phi_j, 
\end{equation}
where $c_j=g_j + i f_j , g_j, f_j \in {\bf Q}$ and $\phi_j$ is a quantum state
which describes the statistical sub-ensemble $S_j$ of $S$ consisting of students
who have read the book $b_j.$ The number $v_j= c_j \bar{c}_j$ gives the frequency of students
$s\in S_j$ in $S,$ i.e., proportional probability $\vert S_j\vert/\vert S\vert$
(where, for a set $O,$ we denote its cardinality by the symbol $\vert O\vert$).

The measurement  ${\cal M}$ is the process of an interaction between
a quantum $I$-transformer (student's brain) and a measurement equipment
(professor's brain). Brains are $I$-systems with very complicated internal structure.
A result of interaction between the brain of a student $s \in S$ and 
the brain of a professor cannot be uniquely determined by the information $b_j.$
Moreover, an attempt to verify the condition $s\in S_j$ (by an additional measurement)
may change the result of the measurement ${\cal M}.$ Therefore 
the property to give the answer $a_k$ as a result of the measurement ${\cal M}$ is not 
 an objective property of elements of the statistical ensemble $S$ described by $\phi.$

{\bf Remark 10.2.} {\small We may use the notion of potentia: each $s\in S$ is potentially 
present in all states $\psi_k=$(the answer $a_k$ to the question $L$). The interaction 
with the equipment induces a transition from possible to actual.}

 The state $\phi$ can be presented in the form:
\begin{equation}
\label{xe1}
\phi = \sum d_k \psi_k, 
\end{equation}
where $d_k =m_k + i n_k, m_k, n_k \in {\bf Q}.$ Probability to obtain
the answer $a_k$ is given by the standard formula
$u_k = d_k \bar{d}_k.$ We could not consider probability 
$u_k$ as probability with respect   to a statistical ensemble. 
This is not proportional probability of the form $\vert S_k^\prime \vert/\vert S\vert,$
where $S_k^\prime$ is a statistical sub-ensemble of $S.$ In fact, the expansion
(\ref{xe1}) provides a description of some properties of an individual system
$s\in S$ (reactions of $s$ to the question $L$). However, we cannot assume that
$\phi$ provides a complete description of the $I$-state of 
a cognitive system $s.$ Thinking systems of students can
be very different\footnote{The complete description of the $I$-states can be obtained
on the basis of hidden variables models. The use of hidden information parameters 
is very natural in quantum information theory.  For example, the brain contains a large
number of information parameters which determine results of $I$-measurements.
These parameters are really hidden in the subconsciousness. The consciousness cannot
control them (see [18] for the details).}.

As usual, we introduce a diagonal operator $A$ in a $p$-adic Hilbert space,
$A \psi_i = a_i \psi_i.$ The spectrum of $A$ coincides with the set of answers ${\cal A}.$
This operator provides the quantum description of the measurement ${\cal M}.$

As we have already noted, the act on the observation is a part of the measurement process
${\cal M}.$ In our example it is important that a student must give an answer to a
professor. If we change the measurement procedure and consider a self-observation
instead of an answer to the professor, then the states $\psi_i$ will be changed 
(with the corresponding change of probabilities).

Finally we have to remark that the quantum $I$-formalism can be used to construct 
a new model for  Bohm's pilot wave theory.  In fact our approach
is quite adequate to ideas of D. Bohm and B. Hiley [67] on active information. 
Moreover, it seems that the ordinary pilot wave theory might be improved
by considering the $\psi$-function field as a purely $I$-field. }

\newpage

{\bf REFERENCES}

1. N. Bohr, {\it Atomic theory and the description of nature.}
Cambridge Univ. Press,
Cambridge(1961).

2. E. Schr\"odinger, {\it My view of the world.} Cambridge Univ. Press,
Cambridge(1964).

3. E. Schr\"odinger, {\it Mind and Matter.} Cambridge Univ. Press,
Cambridge(1958).

4. D. Bohm, Quantum theory as an indications of a new order in physics. 
Part B. Implicate and explicate order in physical law. {\it Found. Phys.},
{\bf 1}, 139-168(1971).

5. E. P. Wiegner, Remark on the mind body problem. In I.J. Good(Ed.),
{\it The scientist speculates.} (p.284-302). Basic Books Inc., 
New York(1962).

6. B. d'Espagnat, {\it Veiled Reality. An analysis of present-day
quantum mechanical concepts.} Addison-Wesley publishing company (1995).

7. B. d'Espagnat, {\it Reality and Physicist.} Cambridge Univ. Press (1989).

8. G. Ryle, {\it The Concept of Mind,\/} Barnes \& Noble, New York (1949).

9. L. E. Rhine, {\it Mind over matter,} Macmillan, New York(1970).

10. D. O. Hebb, {\it Essay on Mind.} Lawrence Erlbaum Associates Publ., Hillsdale,
New Jersey(1980).

11. J. R. Searle, Minds, brains and programs. {\it The Behavioral and Brain
Sciences.} {\bf 3}, 417-457(1980).

12. H. Schmidt, Can an effect precede its case? {\it Found. Phys.},
{\bf 8}, 463-480(1981)

13. H. Schmidt, Callapse of the state vector and psychokinetic effect.
{\it Found. Phys.}, {\bf 12}, 565-581(1982).

14. R.G. Jahn, B.J. Dunne, On the quantum mechanics of consciousness,
with applications to anomalous phenomena. {\it Found. Phys.},
{\bf 16}, 721-772(1986).

15. B. J. Dunne, Y. H. Dobyns, R. G. Jahn, R. D. Nelson,
Series position effects in random event generator experiments.
{\it with an Appendix of } A. Thompson,
Serial position effects in the psychological literature.
{\it J. of Scientific Exploration}, {\bf 8}, 197-215(1994).

16. D. L. Radin, R. D. Nelson, Evidence for consciousness-related anamalies in
random physical systems. {\it Found. Phys.}, {\bf 19}, 1499- 1514
(1989). 

17. H. Smidt, The strange properties of psychokinesis. {\it J. of Scientific
Exploration}, {\bf 1}, No. 1, 1-14(1987).

18. A. Yu. Khrennikov, {\it Non-Archimedean analysis: quantum paradoxes, dynamical
systems and biological models.} Kluwer Academic Publ.,
Dordrecht(1997).

19. A. Yu. Khrennikov, Human subconscious as a $p$-adic dynamical system.
{\it J. Theor. Biol.}, to be published.

20. S. Albeverio, A. Khrennikov, P. E. Kloeden, Human memory as a 
$p$-adic dynamical system. {\it Byosystems}, to be published.

21. W. Schikhov, {\it Ultrametric Calculus}, Cambridge Univ. Press,
Cambridge (1984)

22. I. V.  Volovich,  $p$-adic string. {\it Class. Quant. Grav.},
{\bf 4}, 83--87 (1987).

23. V.S. Vladimirov and I. V. Volovich,  
$p$-adic quantum mechanics. {\it Commun. Math. Phys.}, {\bf 123}, 659--676 (1989).

24. P.G.O. Freud and  M. Olson, Non-Archimedean strings.
{\it Phys. Lett. B}, {\bf 199}, 186-190 (1987).

25. P.G.O. Freund  and  E. Witten, Adelic string amplitudes.
{\it Phys. Lett. B}, {\bf 199}, 191-195 (1987).

26. V. S. Vladimirov, I. V. Volovich, and E. I. Zelenov, 
{\it $p$-adic analysis and  Mathematical Physics}, World 
Scientific Publ., Singapure(1993).

27. Frampton P. H. and Okada Y., $p$-adic string
$N$-point function. {\it Phys. Rev. Lett. B }, {\bf 60}, 484--486 (1988).

28. Aref'eva I. Ya., Dragovich  B., Frampton P. H., Volovich I. V.,
The wave function of the Universe and $p$-adic gravity. {\it Int. J. of 
Modern Phys., A}, {\bf 6}, No 24, 4341--4358 (1991).

29. A. Yu.  Khrennikov, {\it $p$-adic valued distributions in 
mathematical physics.} Kluwer Academic Publishers, Dordrecht (1994).

30. A. Yu. Khrennikov, Mathematical
methods of the non-Ar\-chi\-me\-de\-an physics. 
{\it Uspekhi Mat. Nauk}, {\bf 45}, No. 4, 79--110 (1990).

31. Yu. Manin, New dimensions in geometry. {\it Springer Lecture
Notes in Math.}, {\bf 1111}, 59-101 (1985).

32. B. Dragovic, Adelic wave function of the universe. {\it
Proc. 3rd A. Friedmann Inter. Seminar on Grav. Cosmology,}
St. Peterburg (1995).

33. A. Yu. Khrennikov, $p$-adic quantum-classical analogue of the Heisenberg 
uncertainty relations. {\it Il Nuovo Cimento.} {\bf 112 B}, N. 4, 555-560
(1996)

34. S. Albeverio and  A.Yu. Khrennikov, $p$-adic Hilbert space
representation of quantum systems with an infinite number of degrees of
freedom. {\it Inter. J. Modern Phys.}, {\bf 10}, No.13/14, 1665-1673 (1996).

35. S. Albeverio and A. Yu.  Khrennikov, Representation of the
Weyl group
in spaces of square integrable functions with respect to $p$-adic valued
Gaussian distributions. {\it  J. Phys. A},  {\it 29}, 5515-5527 (1996).

36. R. Cianci and  A.Yu. Khrennikov, $p$-adic numbers and the
renormalization of eigenfunctions in quantum mechanics. {\it Phys. Lett.B},
No. 1/2,109-112 (1994).

37. Khrennikov A.Yu., Ultrametric Hilbert space representation of 
quantum mechanics with a finite exactness.  {\it Found. of 
Physics.}, {\bf 26}, No. 8, 1033--1054 (1996).

38. A. Yu. Khrennikov, $p$-adic stochastic hidden variable model
{\it J. of Math. Physics}, 39, No. 3, 1388-1402(1997).

39. A. Yu. Khrennikov,  p-adic probability interpretation of Bell's inequality paradoxes.
{\it Physics Letters A}, {\bf 200}, 119--223 (1995). 

40. A. Yu. Khrennikov, $p$-adic probability distribution of hidden 
variables. {\it Physica A}, {\bf 215}, 577--587 (1995).

41. J. von Neumann, {\it Mathematical foundations of quantum mechanics.}
Princeton Univ. Press, Princeton(1955).

42. M. Jammer, {\it The conceptual development of quantum mechanics.}
McGraw-Hill, New York(1966).

43. J. M. Jauch, {\it Foundations of quantum mechanics.} Addison-Wesley,
Reading, Mass.(1968).

44. L.E. Ballentine, The statistical interpretation of quantum mechanics.
{\it Rev. Mod. Phys.}, {\bf 42}, 358 - 381(1970).

45. L. E. Ballentine, {\it Quantum mechanics}, Englewood Cliffs, 
New \linebreak Jersey (1989).

46. A. Escassut,  {\it Analytic elements in $p$-adic analysis.} 
World Scientic, Singapore, 1995. 

47. S. Freund, {\it New introductory lectures on psychoanalysis.}
Norton, New York(1933).

48. Kolmogorov A. N., {\it Foundations of the Probability Theory}, 
Chelsea Publ. Comp., New York (1956).

49. Mises R. von, {\it The mathematical theory of probability and
 statistics,} Academic, London(1964).

50. A. Yu. Khrennikov,  $p$-adic theory of probability and its 
applications. A principle of the statistical stabilization of frequencies.
 {\it Teor. and Matem. Fizika}, {\bf 97}, No. 3, 348--363 (1993)

51. Feynman R.P., Negative probability. {\it Quantum Implications,  
Essays in Honour of David Bohm}, B.J. Hiley and F.D. Peat, editors,
Routledge and Kegan Paul, London, 235--246 (1987).

52. Muckenheim W., A review on extended probabilities.
{\it Phys. Reports,} {\bf 133}, 338--401 (1986).

53. Dirac P.A.M., The physical interpretation of quantum mechanics, 
{\it Proc. Roy. Soc. London}, {\bf A 180}, 1--39 (1942).

54. Khrennikov A.Yu., $p$-adic probability description of Dirac`s
hypothetical world. {\it  Int. J. Theor. Phys.}, {\bf 34},
No.12, 2423--2434 (1995).

55. Khrennikov A.Yu., Statistical interpretation of $p$-adic quantum
theories with $p$-adic valued wave functions. {\it J. Math. Phys.}, {\bf 36},
No.12, 6625--6632 (1995).

56. E. B. Davies, {\it Quantum theory of open systems}, Academic Press, 
London (1976).

57.  Holevo A. S., {\it Probabilistic and 
statistical aspects of quantum 
theory.} North-Holland, Amsterdam (1982).

58. Ludwig G., {\it Foundations of quantum mechanics}, Springer, 
Berlin (1983).

59. Schild A., Discrete space-time and integral Lorentz transformations. 
{\it Phys. Rev.,} {\bf 73}, 414-425(1948).

60. Hellund E.J., Tanaka K., Quantized space-time. {\it Phys. Rev.,}
{\bf 94}, 192-208(1954).

61. Ahmavaara Yr., The structure of space and the formalism of relativistic quantum theory.
{\it J. Math. Phys.,} {\bf 6}, 87-93 (1965).

62.  Eccles J. C., {\it The understanding of the brain.} McGraw-Hill Book
Company, New-York(1974).

63. Amit D.J., {\it Modeling of brain functions.} Cambridge University Press,
Cambridge(1989).

64. Cohen J.D., Perlstein W.M., Braver T.S., Nystrom L.E., Noll D.C.,
Jonides J. and E.E. Smith, Temporal dynamics of brain activation
during working memory task. {\it Nature,} {\bf 386}, April 10, 604-608
(1997).

65. Courtney S.M., Ungerleider L.G., Keil K. and Haxby J.V.,
Transient and susteined activity in a disturbed neural system for human
working memory. {\it Nature,} {\bf 386}, April 10, 608-611,
(1997).

66. Hoppensteadt F.C., {\it  An introduction to the mathematics of neurons:
modeling in the frequency domain,} Second Ed., Cambridge University Press,
New York(1997).

67.  Bohm D. and Hiley B. , {\it The undivided universe:
an ontological interpretation of quantum mechanics.} Routledge and Kegan Paul, 
London (1993).

\end{document}